\documentstyle[prd,epsf,aps]{revtex}
\input{psfig}

\newcommand{\jp}{J/\psi}
\begin{document}
\draft
\title{
Observation of diffractive $\jp$ production at the Fermilab Tevatron}
\maketitle

\font\eightit=cmti8
\def\r#1{\ignorespaces $^{#1}$}
\hfilneg
\begin{sloppypar}

\vglue 1em
\centerline{\large The CDF Collaboration}

\vglue 1em
\centerline{\large (Submitted to Physical Review Letters)}
\vglue 1em

\noindent
T.~Affolder,\r {23} H.~Akimoto,\r {45}
A.~Akopian,\r {37} M.~G.~Albrow,\r {11} P.~Amaral,\r 8  
D.~Amidei,\r {25} K.~Anikeev,\r {24} J.~Antos,\r 1 
G.~Apollinari,\r {11} T.~Arisawa,\r {45}
% ...
\end{sloppypar}
\vskip .026in

\vskip .026in
\begin{center}
\r 1  {\eightit Institute of Physics, Academia Sinica, Taipei, Taiwan 11529, 
Republic of China} \\
\r 2  {\eightit Argonne National Laboratory, Argonne, Illinois 60439} \\
\r 3  {\eightit Istituto Nazionale di Fisica Nucleare, University of Bologna,
I-40127 Bologna, Italy} \\
\r 4  {\eightit Brandeis University, Waltham, Massachusetts 02254} \\
\r 5  {\eightit University of California at Davis, Davis, California  95616} \\
\r 6  {\eightit University of California at Los Angeles, Los 
Angeles, California  90024} \\  
\r 7  {\eightit Instituto de Fisica de Cantabria, CSIC-University of Cantabria, 
39005 Santander, Spain} \\
\r 8  {\eightit Enrico Fermi Institute, University of Chicago, Chicago, 
Illinois 60637} \\
\r 9  {\eightit Joint Institute for Nuclear Research, RU-141980 Dubna, Russia}
\\
\r {10} {\eightit Duke University, Durham, North Carolina  27708} \\
\r {11} {\eightit Fermi National Accelerator Laboratory, Batavia, Illinois 
60510} \\
\r {12} {\eightit University of Florida, Gainesville, Florida  32611} \\
\r {13} {\eightit Laboratori Nazionali di Frascati, Istituto Nazionale di Fisica
               Nucleare, I-00044 Frascati, Italy} \\
\r {14} {\eightit University of Geneva, CH-1211 Geneva 4, Switzerland} \\
\r {15} {\eightit Glasgow University, Glasgow G12 8QQ, United Kingdom}\\
\r {16} {\eightit Harvard University, Cambridge, Massachusetts 02138} \\
\r {17} {\eightit Hiroshima University, Higashi-Hiroshima 724, Japan} \\
\r {18} {\eightit University of Illinois, Urbana, Illinois 61801} \\
\r {19} {\eightit The Johns Hopkins University, Baltimore, Maryland 21218} \\
\r {20} {\eightit Institut f\"{u}r Experimentelle Kernphysik, 
Universit\"{a}t Karlsruhe, 76128 Karlsruhe, Germany} \\
\r {21} {\eightit Center for High Energy Physics: Kyungpook National
University, Taegu 702-701; Seoul National University, Seoul 151-742; and
SungKyunKwan University, Suwon 440-746; Korea} \\
\r {22} {\eightit High Energy Accelerator Research Organization (KEK), Tsukuba, 
Ibaraki 305, Japan} \\
\r {23} {\eightit Ernest Orlando Lawrence Berkeley National Laboratory, 
Berkeley, California 94720} \\
\r {24} {\eightit Massachusetts Institute of Technology, Cambridge,
Massachusetts  02139} \\   
\r {25} {\eightit University of Michigan, Ann Arbor, Michigan 48109} \\
\r {26} {\eightit Michigan State University, East Lansing, Michigan  48824} \\
\r {27} {\eightit University of New Mexico, Albuquerque, New Mexico 87131} \\
\r {28} {\eightit The Ohio State University, Columbus, Ohio  43210} \\
\r {29} {\eightit Osaka City University, Osaka 588, Japan} \\
\r {30} {\eightit University of Oxford, Oxford OX1 3RH, United Kingdom} \\
\r {31} {\eightit Universita di Padova, Istituto Nazionale di Fisica 
          Nucleare, Sezione di Padova, I-35131 Padova, Italy} \\
\r {32} {\eightit University of Pennsylvania, Philadelphia, 
        Pennsylvania 19104} \\   
\r {33} {\eightit Istituto Nazionale di Fisica Nucleare, University and Scuola
               Normale Superiore of Pisa, I-56100 Pisa, Italy} \\
\r {34} {\eightit University of Pittsburgh, Pittsburgh, Pennsylvania 15260} \\
\r {35} {\eightit Purdue University, West Lafayette, Indiana 47907} \\
\r {36} {\eightit University of Rochester, Rochester, New York 14627} \\
\r {37} {\eightit Rockefeller University, New York, New York 10021} \\
\r {38} {\eightit Rutgers University, Piscataway, New Jersey 08855} \\
\r {39} {\eightit Texas A\&M University, College Station, Texas 77843} \\
\r {40} {\eightit Texas Tech University, Lubbock, Texas 79409} \\
\r {41} {\eightit Institute of Particle Physics, University of Toronto, Toronto
M5S 1A7, Canada} \\
\r {42} {\eightit Istituto Nazionale di Fisica Nucleare, University of Trieste/
Udine, Italy} \\
\r {43} {\eightit University of Tsukuba, Tsukuba, Ibaraki 305, Japan} \\
\r {44} {\eightit Tufts University, Medford, Massachusetts 02155} \\
\r {45} {\eightit Waseda University, Tokyo 169, Japan} \\
\r {46} {\eightit University of Wisconsin, Madison, Wisconsin 53706} \\
\r {47} {\eightit Yale University, New Haven, Connecticut 06520} \\
\r {(\ast)} {\eightit Now at Carnegie Mellon University, Pittsburgh,
Pennsylvania  15213} \\
\r {(\ast\ast)} {\eightit Now at Northwestern University, Evanston, Illinois 
60208}
\end{center}

\centerline{\large Abstract}

\begin{abstract}
We report the first observation of diffractive $\jp(\rightarrow \mu^+\mu^-)$
production in $\bar pp$ collisions at $\sqrt{s}$=1.8 TeV. Diffractive events
are identified by their rapidity gap signature. In a sample of events with two
muons of transverse momentum $p_T^{\mu}>2$ GeV/$c$ within the pseudorapidity
region $|\eta|<$1.0, the ratio of diffractive to total $J/\psi$ production
rates is found to be $R_{J/\psi}= [1.45\pm 0.25]\%$. The ratio $R_{\jp}(x)$
is presented as a function of $x$-Bjorken. By combining it with our
previously measured corresponding ratio $R_{jj}(x)$ for diffractive dijet
production, we extract a value of $0.59\pm 0.15$ for the gluon fraction of the
diffractive structure function of the proton. 
\end{abstract}

In the course of our studies of high energy
$\bar pp$ interactions at the Fermilab Tevatron
using the Collider Detector at Fermilab (CDF),
we have observed a class of events incorporating both a hard 
(high transverse momentum) partonic
scattering and the characteristic signature of single
diffraction dissociation, namely a leading proton or antiproton and a
forward rapidity gap, defined as the absence of particles in a forward
pseudorapidity ($\eta$)~\cite{pseudo} region.
The rapidity gap in such ``hard diffraction" processes
is attributed to the  exchange
of a Pomeron~\cite{Collins}, which in QCD is
a gluon/quark color-singlet construct with the quantum numbers of the vacuum.
Experiments on hard diffraction can be used to address the
question of whether the Pomeron has a unique,
factorizable partonic structure.

In four previous Letters, we reported
results from diffractive $W$-boson~\cite{CDF_W},
dijet~\cite{CDF_JET}, and
$b$-quark~\cite{CDF_b} production obtained using the rapidity gap signature,
and dijet production in association with a leading
antiproton~\cite{CDF_RP}. These results include measurements of
the gluon fraction~\cite{CDF_JET,CDF_b} and of the $x_{bj}$ (Bjorken $x$) 
dependence of the diffractive structure function of the 
antiproton~\cite{CDF_RP}.
Comparisons with diffractive deep inelastic scattering
data obtained at the DESY $ep$ collider HERA revealed a severe
breakdown of QCD factorization, expressed mainly as
a suppression of a factor of $\sim 10$ of the overall normalization of
the diffractive structure function at the Tevatron.  
A breakdown of factorization was also observed between 
the diffractive structure functions measured from dijet production in  
single diffraction and in double Pomeron exchange
at the Tevatron~\cite{CDF_DPE}.
In this Letter, we report a measurement of
diffractive $\jp$ production
in $\bar pp$ collisions at $\sqrt{s}=$1800 GeV,
$p\bar p\rightarrow p({\rm or}\;\bar p)+\jp+X$,
and compare the $J/\psi$ diffractive fraction
with the results of our previous hard diffraction
measurements to further characterize the observed
breakdown of factorization.

The data used in this analysis were collected
during 1994-95 and correspond to an integrated luminosity 
of 80~pb$^{-1}$.
The technique we use to extract the diffractive signal is 
identical to that used in our previous studies.
In a data sample satisfying selection requirements for $\jp$ 
decaying into $\mu^+\mu^-$, we look for events
with a rapidity gap in either of the two forward regions 
of the detector covering the pseudorapidity range $2.4<|\eta|<5.9$.
We define a rapidity gap as the absence of hits in the beam-beam 
counters (BBC), 
which cover the region 3.2$<|\eta|<$5.9, and the absence of calorimeter 
towers with energy above 1.5 GeV within $2.4<|\eta|<4.2$.
The size of the calorimeter towers in this region is
$\Delta\eta\times \Delta\phi=0.1\times 5^{\circ}$.

The detector components relevant to $\jp$ selection 
are the silicon vertex detector (SVX), the vertex time projection chamber 
(VTX), the central tracking chamber (CTC), the central electromagnetic and 
hadronic calorimeters surrounding the CTC, and the central muon detectors.
The $\jp$ acceptance is limited by the muon detectors, which cover the region 
$|\eta|<1.0$.
The SVX provides spatial measurements in the $r$-$\phi$ plane with a track 
impact parameter resolution of $[13+(40\,{\rm GeV}/c)/p_T]\;\mu$m. 
The VTX is used primarily to measure the longitudinal position $z$ 
of an event's primary vertex, and the CTC provides momentum analysis 
for charged particles. The combined CTC/SVX transverse momentum 
resolution for charged 
particles is $\sigma_{p_T}/p_T=0.0009p_T\oplus 0.0066$, 
where $p_T$ is in GeV/$c$.
Details of the CDF detector components can be found in Ref.~\cite{CDF}.

A three-level dimuon trigger system was used to select events with a pair of 
muon candidates~\cite{Thesis}. 
At Level 1, the dimuon trigger selection required the
presence of two radially aligned pairs of time-correlated hits
in the muon chambers.
Level 2 required that each pair of muon chamber hits match a track
in the CTC found by the Central Fast Track 
(CFT) processor, which performed a partial reconstruction of all CTC 
tracks and determined the $p_T$ with a momentum resolution of 
$\sigma _{p_T}/p_T^2=0.035\,({\rm GeV}/c)^{-1}$.
At Level 3, performed in software, events were required
to contain two oppositely charged muon candidates with an invariant mass
within $300$~MeV$/c^2$ of the $J/\psi$ mass of 
$3096.9$~MeV$/c^2$~\cite{jpsi_mass}.
  
In addition to the $J/\psi$ selection requirements used 
in previous CDF analyses~\cite{PRD_ct}, the following two requirements    
were imposed on the data: 
first, since the BBC information is used to tag rapidity gaps,
only data for which there was no BBC coincidence requirement in the trigger 
were considered; and second, since additional interactions in 
the same beam-beam crossing would most likely spoil a diffractive rapidity 
gap, only events with one reconstructed primary vertex were retained.
In order to ensure that reconstructed muons were found in the
kinematic region where the trigger is highly efficient,
a minimum transverse momentum of $2$~GeV/$c$ was required for each muon 
candidate.
For a precise vertex measurement, both muons were required to be
reconstructed in the SVX detector.
The dimuon invariant mass distribution for events passing the above 
requirements is shown in Figure~1a.
The signal region, defined as the dimuon mass range of 
$3.05\leq M_{\mu^{+}\mu^{-}}< 3.15$ GeV/$c^{2}$, 
contains 18910 events.

There are three sources of dimuons in the above $J/\psi$ 
candidate event sample:
  (a) $J/\psi$'s directly produced in $p\bar{p}$ collisions, 
      or resulting from
      decays of intermediate states which are sufficiently short-lived 
      ($e.g.$ from $\chi_{c}$ decays) so that
      their measured decay vertex cannot be distinguished from the primary 
      event vertex;
  (b) $J/\psi$'s from $B$-hadron decays, and
  (c) background from processes for which the dimuon invariant mass 
      falls accidentally
      in the $J/\psi$ signal mass window; 
      the latter includes dimuons from Drell-Yan
      production, double semileptonic $b$-decays, meson decays-in-flight, and 
      residual hadrons that penetrate the calorimeter and are 
      misidentified as muons.
The first two sources contribute to the $J/\psi$ invariant mass peak, while 
the last comprises the background under the peak.
The fraction of background events in the signal region is evaluated by fitting 
the dimuon mass distribution with a sum of a Gaussian and a linear
function. The fit yields a background fraction of $(6.5\pm0.1)\%$ within the 
signal region.
 
The fraction of $J/\psi$'s from  $B$-hadron decays can be 
determined by fitting  the proper decay length  distribution, $c\tau$, 
using the appropriate function for 
each of the three dimuon components described above.
However, because we do not fully reconstruct $B$ decays, we use an
approximation to $c\tau$ described in~\cite{PRD_ct} and referred to as 
pseudo-$c\tau$.
In the signal region, the fraction of background events is fixed at 
the value of $0.065$, obtained from the dimuon mass fit,
and the pseudo-$c\tau$ distribution is fitted using for the background a
parametrization derived from the sidebands and appropriate 
parametrizations for the prompt and $B$ decay dimuon 
components~\cite{Thesis,PRD_ct}.
The pseudo-$c\tau$ distribution for the signal region is shown in Fig.~1b 
with the fit result superimposed. The fraction of 
$J/\psi$ mesons from $B$-hadron decays obtained 
from the fit is $(16.8\pm0.4)\%$.
The vertical line at $100~\mu$m separates two regions: a ``long-lived" region
dominated by $B$ decays,  and a ``short-lived" region 
mostly due to prompt $J/\psi$ mesons. The short-lived region contains
15824 events, which are used in the analysis below.
By numerically integrating the fitted $B$ decay component in this region, 
the $B$-hadron decay contamination is found to be $3.3\%$.

As in our previous rapidity gap 
studies~\cite{CDF_W,CDF_JET,CDF_b}, the diffractive signal 
is evaluated by considering the number of BBC hits, $N_{\rm BBC}$, 
versus the number of the adjacent forward calorimeter towers with 
energy above 1.5 GeV, $N_{\rm CAL}$. 
Figure 2a shows the correlation between $N_{\rm BBC}$ and $N_{\rm CAL}$.
The multiplicity in this figure is for the side of the detector with 
the lower BBC hit multiplicity.
The (0,0) bin, $N_{\rm BBC}=N_{\rm CAL}=0$, contains 92 events.
The excess of events in this bin is attributed to diffractive production.
The non-diffractive content of the (0,0) bin is evaluated from the diagonal of
Fig.~2a with $N_{\rm BBC}=N_{\rm CAL}$, shown in Fig.~2b.
The non-$J/\psi$ background in each bin of this plot, estimated by
fitting the dimuon mass distribution to the sum of
a Gaussian and a constant function, was subtracted from the number of $J/\psi$
candidates prior to plotting, yielding $87.4\pm 9.7$ $J/\psi$ events 
in the (0,0) bin. 
An extrapolation to bin (0,0) of a linear fit to the data of bins (2,2) to 
(12,12) yields $19.9\pm3.9$ non-diffractive events.
The events in the (0,0) bin will be referred to as 
``diffractive''. 
Figures~2c and 2d show the $J/\psi$ transverse momentum 
and pseudorapidity distribution, respectively, for the diffractive and
total event samples. In Fig.~2d the sign of the $J/\psi$ pseudorapidity 
for events with a gap at positive $\eta$ is reversed, so that the gap 
appears always on the left.

The number of diffractive events in the (0,0) bin must be corrected for 
the efficiency of requiring a single reconstructed primary vertex, 
$\varepsilon_{\rm 1vtx}^{\rm SD}$, as well 
as for random BBC and forward calorimeter occupancy.
The single-vertex requirement, which is used to reject events due to 
multiple interactions, also rejects single interaction events with extra
vertices due to track reconstruction ambiguities.
Removing the single-vertex requirement yields 15.4 more 
diffractive $J/\psi$ events, resulting in 
$\varepsilon_{\rm 1vtx}^{\rm SD}=0.85$.
For non-diffractive events, the efficiency of the single-vertex requirement, 
$\varepsilon_{\rm 1vtx}^{\rm ND}$, was evaluated by comparing
the ratio of single-vertex to all events with the 
ratio expected from the instantaneous luminosity. This comparison yielded 
$\varepsilon_{\rm 1vtx}^{\rm ND}=0.56\pm 0.04$.
Finally, from a study of a sample of events with no reconstructed 
primary vertex collected in random beam-beam crossings, the combined 
BBC and forward calorimeter occupancy was measured to be $0.20\pm0.06$.

After correcting the data for the efficiency of the single-vertex 
requirement and for the forward detector occupancy, 
we obtain a diffractive to total $J/\psi$ 
production ratio of $R_{J/\psi}^{\rm gap}=(0.42\pm0.07)\%$.
This ratio is based on diffractive events satisfying our rapidity gap
definition. Therefore, it must be corrected for the 
rapidity gap acceptance, $\varepsilon^{gap}$, defined as the ratio of 
events in bin (0,0) to the total 
number of diffractive events satisfying the same $J/\psi$ requirements  
and produced within a specified range of $\xi$, 
where $\xi$ is the fractional momentum loss of the leading (anti)proton.
The gap acceptance for $\xi<0.1$ was
calculated using the POMPYT Monte Carlo
generator~\cite{POMPYT} followed by a detector simulation.
For a Pomeron structure function of the form 
${\beta}f(\beta)=1/\beta$~\cite{CDF_RP}, where $\beta$
is the fraction of the momentum of the Pomeron carried by a parton,
$\varepsilon^{gap}$ was found to be $0.29$.
Dividing $R_{J/\psi}^{\rm gap}$ by $\varepsilon^{gap}$ 
yields a diffractive to total production ratio of
$R_{J/\psi}=(1.45\pm0.25)\%$.
    
The ratio $R_{J/\psi}$ is larger than the corresponding ratio for 
diffractive $b$-quark production, $R_{\bar bb}=(0.62\pm 0.25)\%$~\cite{CDF_b}, 
by a factor of $2.34\pm0.35$.  
As both $J/\psi$ and $b$-quark production are mainly sensitive to the 
gluon content of the Pomeron, we examine whether the difference in the 
two ratios could be attributed to the different average $x_{bj}$ values 
of the two measurements. Given the $x_{bj}^{-0.45}$ dependence of the 
diffractive structure function measured in dijet production~\cite{CDF_RP}, 
the double ratio $R_{\bar bb}^{J/\psi}\equiv R_{J/\psi}/R_{\bar bb}$
is expected to be equal to $(x_{bj}^{J/\psi}/x_{bj}^{\bar bb})^{-0.45}$.
Since in these measurements we consider only 
central $J/\psi$ or $b$-quark production, the ratio 
$x_{bj}^{J/\psi}/x_{bj}^{\bar bb}$ is approximately proportional to the 
ratio of the corresponding average $p_T$ value for each process, which is 
$\approx 6$ GeV/$c$ for the $J/\psi$ (see Fig.~2c) and $\approx 36$ GeV/$c$ for 
the $b$-quark (about three 
times the average $p_T$ of the $b$-decay electron~\cite{CDF_b}). 
The expected value 
for $R_{\bar bb}^{J/\psi}$ is then $\approx (6/36)^{-0.45}=2.2$, 
in agreement with the 
measured value of $2.34\pm0.35$.

For a more direct study of the diffractive structure function, we restricted 
our analysis to events in which at least one jet was reconstructed. 
A jet is defined as a cluster of calorimeter towers within a cone size of
${\Delta}R\equiv(\Delta\eta^2+\Delta\phi^2)^{\frac 12}=0.7$ 
with a seed tower of $E_T>1$~GeV.
Since our diffractive $J/\psi$ events have a rapidity gap in the region
$2.4\le|\eta|\le{5.9}$, the core of the reconstructed jet for both 
diffractive and non-diffractive
events is restricted to the region $|\eta|<1.7$.
The number of events passing this requirement is 8732.

Figure 3 shows distributions for the $J/\psi+jet$ event sample:
(a) is the diagonal $(N_{\rm CAL},N_{\rm BBC})$ distribution, 
equivalent to that of
Fig.~1b, (b) the corrected $\xi_{p,\bar p}$ 
distribution for the (anti)proton on the side of the gap, evaluated 
using calorimeter and BBC information in a procedure described in 
Ref.~\cite{CDF_DPE},
(c) the $J/\psi$ transverse momentum, and (d) the
azimuthal angle difference, $\phi=|\phi_{J/\psi}-\phi_{jet}|$,
between the $J/\psi$ and the highest $E_T$ jet.   

The $x_{bj}$ of the parton in the (anti)proton participating in 
$J/\psi$ production is evaluated using the equation
$x_{bj}^\pm=p_T^{J/\psi}(e^{\pm \eta_{J/\psi}}+e^{\pm \eta_{jet}})/\sqrt{s}$,
where the + ($-$) sign stands for $p(\bar p)$.
In leading order QCD calculations, 
the ratio of diffractive to non-diffractive production 
is equal to the ratio of the corresponding structure functions. 
For $J/\psi$ production, the ratio $R_{J/\psi}(x)$ per unit $\xi$ was 
evaluated for the events in the region $0.01<\xi<0.03$ (see Fig.~3b)
and is plotted in Fig.~4 along with the same ratio for dijet 
production, $R_{jj}(x)$, obtained 
from Ref.~\cite{CDF_RP}. The structure function 
relevant to dijet production is 
$F_{jj}(x)\sim g(x)+\frac{4}{9}q(x)$~\cite{CDF_RP},
where $g(x)$ and $q(x)$ are the gluon and quark densities in the proton
and $\frac 49$ is a color factor. 
For $J/\psi$ production, which  is dominated by $gg$ interactions, 
$R_{J/\psi}(x)\simeq {g^D(x)}/{g(x)}$.
The ratio of $R_{jj}(x)$ to $R_{J/\psi}(x)$ is given by 
\begin{equation}
R^{jj}_{J/\psi}(x)\equiv
\frac{R_{jj}(x)}{R_{J/\psi}(x)}=\frac{1+\frac{4}{9}\frac{q^D(x)}
 {g^D(x)}}{1+\frac{4}{9}\frac{q(x)}{g(x)}}
\label{eq:r-x}
\end{equation}
where the superscript $D$ is used to label the diffractive parton densities.
Evaluating this ratio of ratios by integrating 
the $x_{bj}$ distributions for
$R_{jj}$ and $R_{J/\psi}$ in the region $0.004\le x\le0.01$ (kinematic
boundaries for full acceptance) yields
$[R_{jj}(x)/R_{J/\psi}(x)]_{\rm exp}=1.17\pm0.27({\rm stat})$.
 Using this value in Eq.~\ref{eq:r-x} and the ratio of $q(x)/g(x)=0.274$ 
at $x=0.0063$ and $Q=6$~GeV calculated from the proton GRV98LO 
parton distribution functions~\cite{GRV98LO},
the gluon fraction of the diffractive structure function 
 of the (anti)proton is found 
to be $f_g^D=0.59\pm0.14({\rm stat})\pm0.06({\rm syst})$,
where the systematic uncertainty 
includes in quadrature the uncertainties of all correction factors.
 This gluon fraction is consistent with 
the value $0.54\pm0.15$ obtained by
combining the results of diffractive $W$, dijet, and $b$-quark 
production~\cite{CDF_b}.

     We thank the Fermilab staff and the technical staffs of the
participating institutions for their vital contributions.  This work was
supported by the U.S. Department of Energy and National Science Foundation;
the Italian Istituto Nazionale di Fisica Nucleare; the Ministry of Education,
Science, Sports and Culture of Japan; the Natural Sciences and Engineering
Research Council of Canada; the National Science Council of the Republic of
China; the Swiss National Science Foundation; the A. P. Sloan Foundation; the
Bundesministerium fuer Bildung und Forschung, Germany; the Korea Science
and Engineering Foundation (KoSEF); the Korea Research Foundation; the
Comision Interministerial de Ciencia y Tecnologia, Spain; 
the Max Kade Foundation; and the Ministry of Education, Science and Research
of the Federal State Nordrhein-Westfalen of Germany.

\setcounter{figure}{0}
\vglue 1in
\begin{figure}
\centerline{\psfig{figure=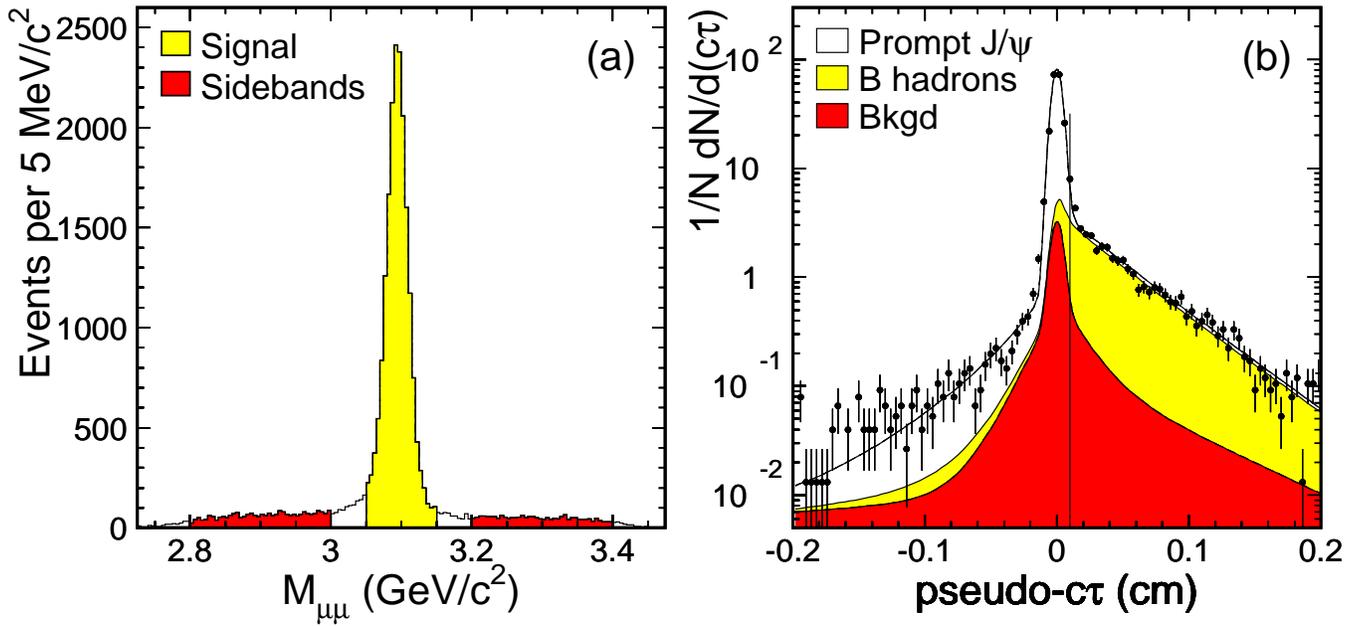,width=7in}}
\vglue 0.5in
\caption{(a) Dimuon invariant mass, and
(b) pseudo-$c\tau$ for signal region with fit to the sum of contributions
from prompt $J/\psi$ production, $J/\psi$ mesons
from $B$-hadron decays and non-$J/\psi$
background.}
\end{figure}
\newpage
\vglue 0.75in
\begin{figure}
\centerline{\psfig{figure=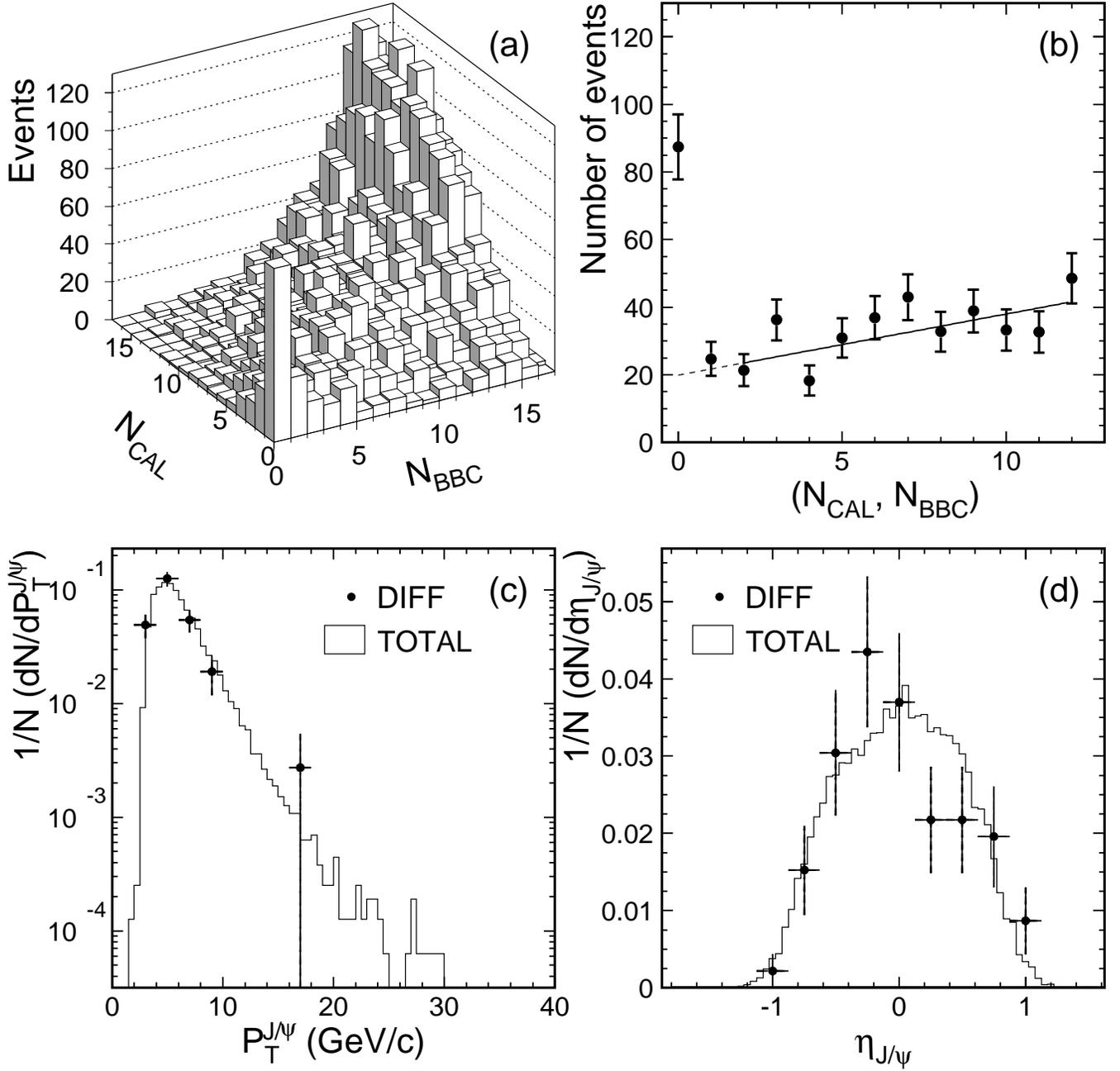,width=7in}}
\vfill
\caption{$J/\psi$ event sample distributions:
(a) Beam-beam counter multiplicity, $N_{\rm BBC}$, for the BBC with the 
lower multiplicity, versus 
forward calorimeter tower multiplicity, $N_{\rm CAL}$;
(b) multiplicity distribution along the diagonal  
with $N_{\rm BBC}=N_{\rm CAL}$ in the plot of (a);
(c) $J/\psi$ transverse momentum and (d) $J/\psi$ pseudorapidity.}
\end{figure}
\newpage
\vglue 0.75in
\begin{figure}
\centerline{\psfig{figure=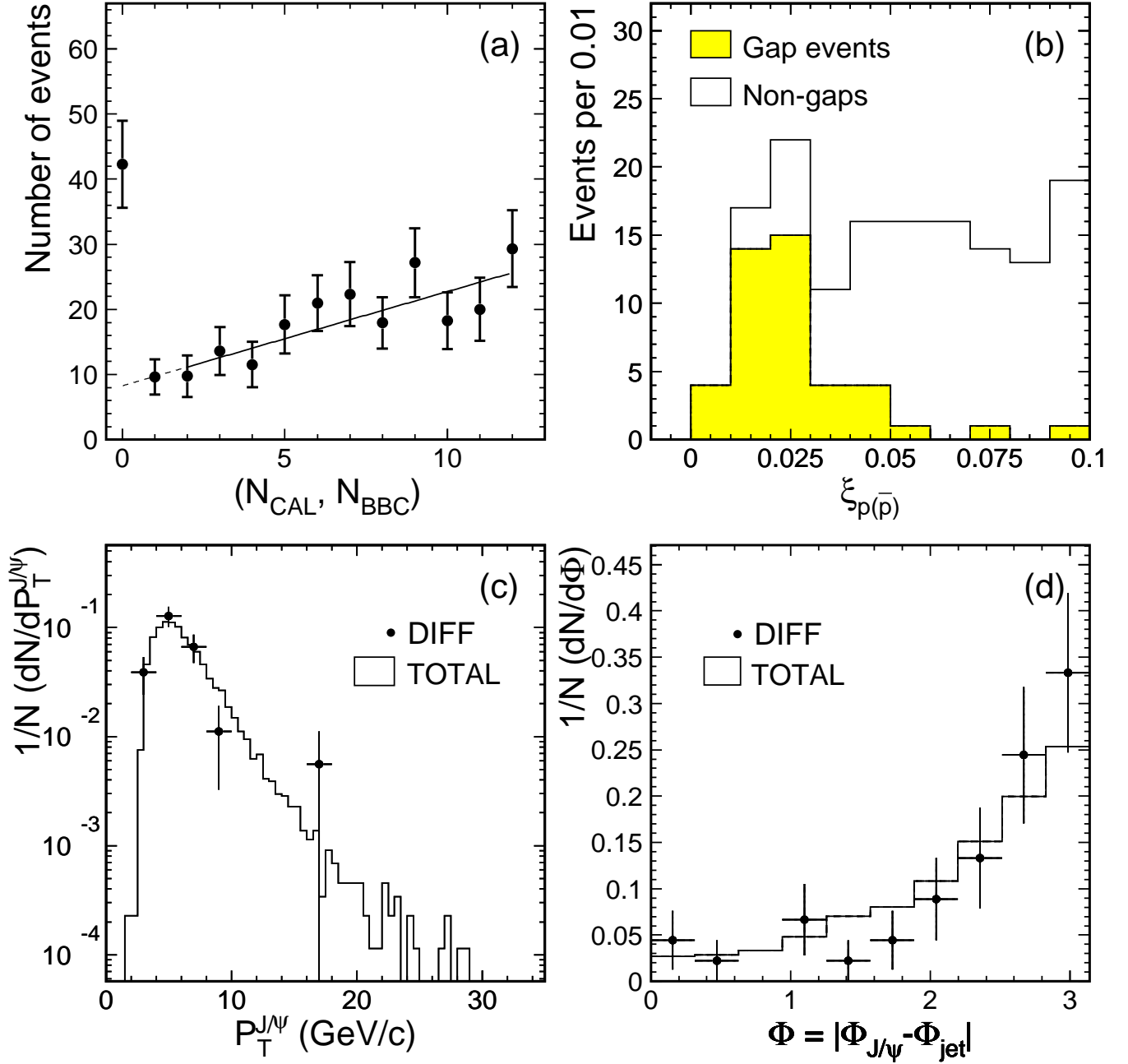,width=7in}}
\vfill
\caption{Distributions for the $J/\psi+jet$ event sample:
(a) the diagonal of the $N_{\rm BBC}$ versus $N_{\rm CAL}$ 
distribution with $N_{\rm BBC}=N_{\rm CAL}$; 
(b) Pomeron beam momentum fraction, 
$\xi_{p,\bar p}$ (corrected), 
for events with the $J/\psi$ within $|\eta|<1.1$
(the shaded area is the distribution for 
events satisfying the rapidity gap requirements);
(c) $J/\psi$ transverse momentum;  (d) azimuthal angle difference between 
the $J/\psi$ and the leading jet.}
\end{figure}
\newpage
\vglue 0.75in
\begin{figure}
\centerline{\psfig{figure=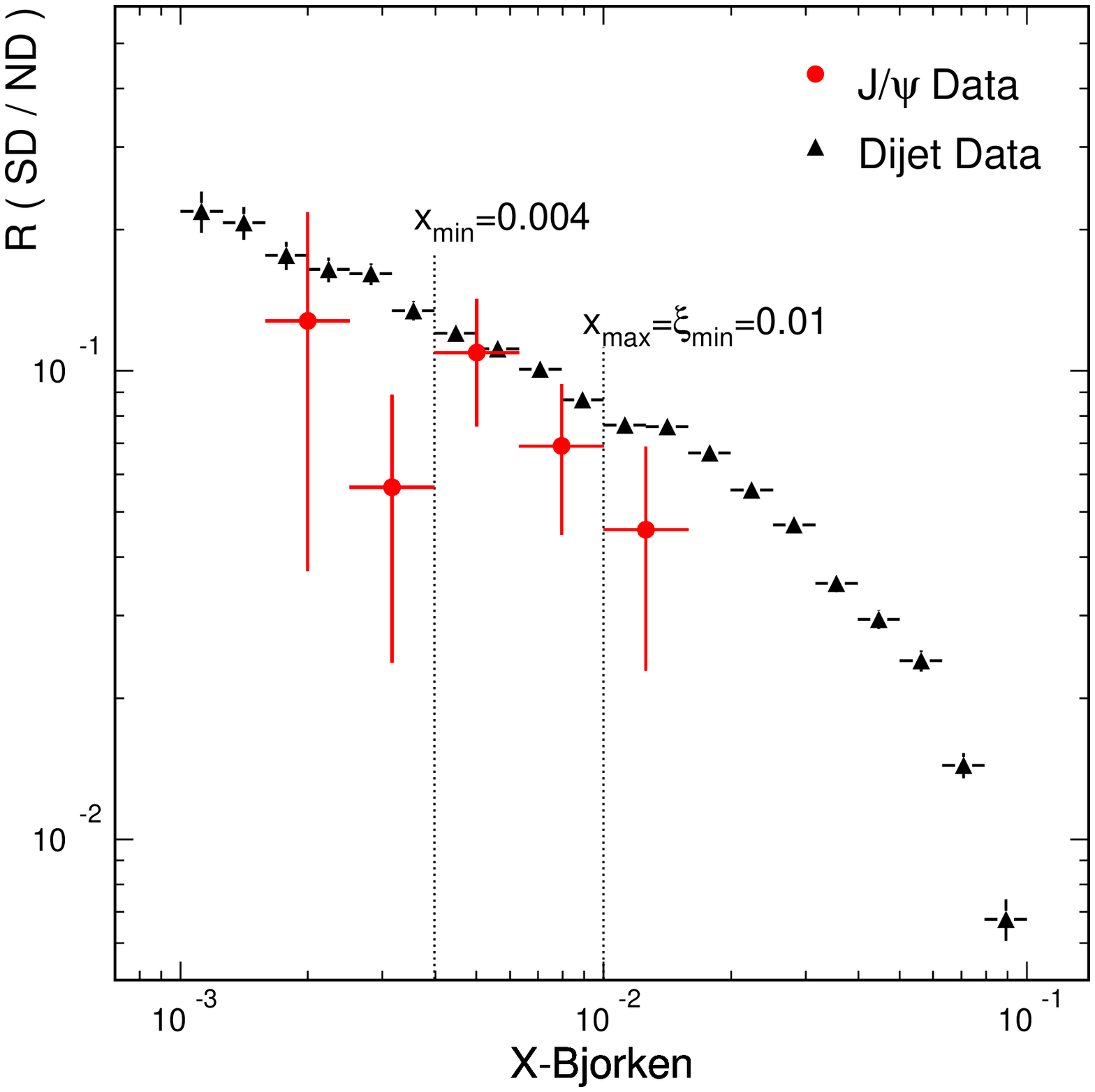,width=7in}}
\vfill
\caption{Ratios of diffractive to non-diffractive $J/\psi$ (circles) and 
dijet (triangles) rates per unit $\xi_p(\xi_{\bar{p}})$ 
as a function of $x$-Bjorken of the struck parton of the 
$p(\bar{p})$ adjacent to the rapidity gap.} 
\end{figure}

\begin{references}
\bibitem{pseudo}  We use rapidity and pseudorapidity, $\eta$, interchangeably; 
		  $\eta\equiv -\ln(\mbox{tan}\frac{\theta}{2})$, 
		  where $\theta$ is the polar angle of a particle with respect 
	          to the proton beam direction.
                  The azimuthal angle is denoted by $\phi$, and transverse 
	          energy is defined as $E_T=E\sin\theta$.
\end{references}
\end{document}